\documentclass[twoside,reqno]{bjp}
\usepackage{epsfig,cite}
\usepackage{graphics}
\usepackage{amssymb,amsmath}
\usepackage{times}
\begin{document}

\sloppy \raggedbottom
\setcounter{page}{1}


%
%
%
%
\newcommand\rf[1]{(\ref{eq:#1})}
\newcommand\lab[1]{\label{eq:#1}}
\newcommand\nonu{\nonumber}
\newcommand\br{\begin{eqnarray}}
\newcommand\er{\end{eqnarray}}
\newcommand\be{\begin{equation}}
\newcommand\ee{\end{equation}}
\newcommand\eq{\!\!\!\! &=& \!\!\!\! }
\newcommand\foot[1]{\footnotemark\footnotetext{#1}}
\newcommand\lb{\lbrack}
\newcommand\rb{\rbrack}
\newcommand\llangle{\left\langle}
\newcommand\rrangle{\right\rangle}
\newcommand\blangle{\Bigl\langle}
\newcommand\brangle{\Bigr\rangle}
\newcommand\llb{\left\lbrack}
\newcommand\rrb{\right\rbrack}
\newcommand\Blb{\Bigl\lbrack}
\newcommand\Brb{\Bigr\rbrack}
\newcommand\lcurl{\left\{}
\newcommand\rcurl{\right\}}
\renewcommand\({\left(}
\renewcommand\){\right)}
\renewcommand\v{\vert}                     
\newcommand\bv{\bigm\vert}               
\newcommand\Bgv{\;\Bigg\vert}            
\newcommand\bgv{\bigg\vert}              
\newcommand\lskip{\vskip\baselineskip\vskip-\parskip\noindent}
\newcommand\mskp{\par\vskip 0.3cm \par\noindent}
\newcommand\sskp{\par\vskip 0.15cm \par\noindent}
\newcommand\bc{\begin{center}}
\newcommand\ec{\end{center}}
\newcommand\Lbf[1]{{\Large \textbf{{#1}}}}
\newcommand\lbf[1]{{\large \textbf{{#1}}}}




\newcommand\tr{\mathop{\mathrm tr}}                  
\newcommand\Tr{\mathop{\mathrm Tr}}                  
\newcommand\partder[2]{\frac{{\partial {#1}}}{{\partial {#2}}}}
\newcommand\partderd[2]{{{\partial^2 {#1}}\over{{\partial {#2}}^2}}}
\newcommand\partderh[3]{{{\partial^{#3} {#1}}\over{{\partial {#2}}^{#3}}}}
\newcommand\partderm[3]{{{\partial^2 {#1}}\over{\partial {#2} \partial{#3} }}}
\newcommand\partderM[6]{{{\partial^{#2} {#1}}\over{{\partial {#3}}^{#4}{\partial {#5}}^{#6} }}}          
\newcommand\funcder[2]{{{\delta {#1}}\over{\delta {#2}}}}
\newcommand\Bil[2]{\Bigl\langle {#1} \Bigg\vert {#2} \Bigr\rangle}  
\newcommand\bil[2]{\left\langle {#1} \bigg\vert {#2} \right\rangle} 
\newcommand\me[2]{\left\langle {#1}\right|\left. {#2} \right\rangle} 

\newcommand\sbr[2]{\left\lbrack\,{#1}\, ,\,{#2}\,\right\rbrack} 
\newcommand\Sbr[2]{\Bigl\lbrack\,{#1}\, ,\,{#2}\,\Bigr\rbrack}
\newcommand\Gbr[2]{\Bigl\lbrack\,{#1}\, ,\,{#2}\,\Bigr\} }
\newcommand\pbr[2]{\{\,{#1}\, ,\,{#2}\,\}}       
\newcommand\Pbr[2]{\Bigl\{ \,{#1}\, ,\,{#2}\,\Bigr\}}  
\newcommand\pbbr[2]{\lcurl\,{#1}\, ,\,{#2}\,\rcurl}




\renewcommand\a{\alpha}
\renewcommand\b{\beta}
\renewcommand\c{\chi}
\renewcommand\d{\delta}
\newcommand\D{\Delta}
\newcommand\eps{\epsilon}
\newcommand\vareps{\varepsilon}
\newcommand\g{\gamma}
\newcommand\G{\Gamma}
\newcommand\grad{\nabla}
\newcommand\h{\frac{1}{2}}
\renewcommand\k{\kappa}
\renewcommand\l{\lambda}
\renewcommand\L{\Lambda}
\newcommand\m{\mu}
\newcommand\n{\nu}
\newcommand\om{\omega}
\renewcommand\O{\Omega}
\newcommand\p{\phi}
\newcommand\vp{\varphi}
\renewcommand\P{\Phi}
\newcommand\pa{\partial}
\newcommand\tpa{{\tilde \partial}}
\newcommand\bpa{{\bar \partial}}
\newcommand\pr{\prime}
\newcommand\ra{\rightarrow}
\newcommand\lra{\longrightarrow}
\renewcommand\r{\rho}
\newcommand\s{\sigma}
\renewcommand\S{\Sigma}
\renewcommand\t{\tau}
\renewcommand\th{\theta}
\newcommand\bth{{\bar \theta}}
\newcommand\Th{\Theta}
\newcommand\z{\zeta}
\newcommand\ti{\tilde}
\newcommand\wti{\widetilde}
\newcommand\twomat[4]{\left(\begin{array}{cc}  
{#1} & {#2} \\ {#3} & {#4} \end{array} \right)}
\newcommand\threemat[9]{\left(\begin{array}{ccc}  
{#1} & {#2} & {#3}\\ {#4} & {#5} & {#6}\\
{#7} & {#8} & {#9} \end{array} \right)}


\newcommand\cA{{\mathcal A}}
\newcommand\cB{{\mathcal B}}
\newcommand\cC{{\mathcal C}}
\newcommand\cD{{\mathcal D}}
\newcommand\cE{{\mathcal E}}
\newcommand\cF{{\mathcal F}}
\newcommand\cG{{\mathcal G}}
\newcommand\cH{{\mathcal H}}
\newcommand\cI{{\mathcal I}}
\newcommand\cJ{{\mathcal J}}
\newcommand\cK{{\mathcal K}}
\newcommand\cL{{\mathcal L}}
\newcommand\cM{{\mathcal M}}
\newcommand\cN{{\mathcal N}}
\newcommand\cO{{\mathcal O}}
\newcommand\cP{{\mathcal P}}
\newcommand\cQ{{\mathcal Q}}
\newcommand\cR{{\mathcal R}}
\newcommand\cS{{\mathcal S}}
\newcommand\cT{{\mathcal T}}
\newcommand\cU{{\mathcal U}}
\newcommand\cV{{\mathcal V}}
\newcommand\cX{{\mathcal X}}
\newcommand\cW{{\mathcal W}}
\newcommand\cY{{\mathcal Y}}
\newcommand\cZ{{\mathcal Z}}

\newcommand{\nit}{\noindent}
\newcommand{\ct}[1]{\cite{#1}}
\newcommand{\bib}[1]{\bibitem{#1}}

\newcommand\PRL[3]{\textsl{Phys. Rev. Lett.} \textbf{#1} (#2) #3}
\newcommand\NPB[3]{\textsl{Nucl. Phys.} \textbf{B#1} (#2) #3}
\newcommand\NPBFS[4]{\textsl{Nucl. Phys.} \textbf{B#2} [FS#1] (#3) #4}
\newcommand\CMP[3]{\textsl{Commun. Math. Phys.} \textbf{#1} (#2) #3}
\newcommand\PRD[3]{\textsl{Phys. Rev.} \textbf{D#1} (#2) #3}
\newcommand\PLA[3]{\textsl{Phys. Lett.} \textbf{#1A} (#2) #3}
\newcommand\PLB[3]{\textsl{Phys. Lett.} \textbf{#1B} (#2) #3}
\newcommand\CQG[3]{\textsl{Class. Quantum Grav.} \textbf{#1} (#2) #3}
\newcommand\JMP[3]{\textsl{J. Math. Phys.} \textbf{#1} (#2) #3}
\newcommand\PTP[3]{\textsl{Prog. Theor. Phys.} \textbf{#1} (#2) #3}
\newcommand\SPTP[3]{\textsl{Suppl. Prog. Theor. Phys.} \textbf{#1} (#2) #3}
\newcommand\AoP[3]{\textsl{Ann. of Phys.} \textbf{#1} (#2) #3}
\newcommand\RMP[3]{\textsl{Rev. Mod. Phys.} \textbf{#1} (#2) #3}
\newcommand\PR[3]{\textsl{Phys. Reports} \textbf{#1} (#2) #3}
\newcommand\FAP[3]{\textsl{Funkt. Anal. Prilozheniya} \textbf{#1} (#2) #3}
\newcommand\FAaIA[3]{\textsl{Funct. Anal. Appl.} \textbf{#1} (#2) #3}
\newcommand\TAMS[3]{\textsl{Trans. Am. Math. Soc.} \textbf{#1} (#2) #3}
\newcommand\InvM[3]{\textsl{Invent. Math.} \textbf{#1} (#2) #3}
\newcommand\AdM[3]{\textsl{Advances in Math.} \textbf{#1} (#2) #3}
\newcommand\PNAS[3]{\textsl{Proc. Natl. Acad. Sci. USA} \textbf{#1} (#2) #3}
\newcommand\LMP[3]{\textsl{Letters in Math. Phys.} \textbf{#1} (#2) #3}
\newcommand\IJMPA[3]{\textsl{Int. J. Mod. Phys.} \textbf{A#1} (#2) #3}
\newcommand\IJMPD[3]{\textsl{Int. J. Mod. Phys.} \textbf{D#1} (#2) #3}
\newcommand\TMP[3]{\textsl{Theor. Math. Phys.} \textbf{#1} (#2) #3}
\newcommand\JPA[3]{\textsl{J. Physics} \textbf{A#1} (#2) #3}
\newcommand\JSM[3]{\textsl{J. Soviet Math.} \textbf{#1} (#2) #3}
\newcommand\MPLA[3]{\textsl{Mod. Phys. Lett.} \textbf{A#1} (#2) #3}
\newcommand\JETP[3]{\textsl{Sov. Phys. JETP} \textbf{#1} (#2) #3}
\newcommand\JETPL[3]{\textsl{ Sov. Phys. JETP Lett.} \textbf{#1} (#2) #3}
\newcommand\PHSA[3]{\textsl{Physica} \textbf{A#1} (#2) #3}
\newcommand\PHSD[3]{\textsl{Physica} \textbf{D#1} (#2) #3}
\newcommand\JPSJ[3]{\textsl{J. Phys. Soc. Jpn.} \textbf{#1} (#2) #3}
\newcommand\JGP[3]{\textsl{J. Geom. Phys.} \textbf{#1} (#2) #3}

\newcommand\Xdot{\stackrel{.}{X}}
\newcommand\xdot{\stackrel{.}{x}}
\newcommand\ydot{\stackrel{.}{y}}
\newcommand\yddot{\stackrel{..}{y}}
\newcommand\rdot{\stackrel{.}{r}}
\newcommand\rddot{\stackrel{..}{r}}
\newcommand\vpdot{\stackrel{.}{\varphi}}
\newcommand\vpddot{\stackrel{..}{\varphi}}
\newcommand\tdot{\stackrel{.}{t}}
\newcommand\zdot{\stackrel{.}{z}}
\newcommand\etadot{\stackrel{.}{\eta}}
\newcommand\udot{\stackrel{.}{u}}
\newcommand\vdot{\stackrel{.}{v}}
\newcommand\rhodot{\stackrel{.}{\rho}}
\newcommand\xdotdot{\stackrel{..}{x}}
\newcommand\ydotdot{\stackrel{..}{y}}


\title{A New Mechanism of Dynamical Spontaneous Breaking of Supersymmetry
\thanks{Invited talk at the International Conference {\em ``Mathematics Days in Sofia''}, 
July 2014, \textsl{http://www.math.bas.bg/mds2014/}}}

\begin{start}
\author{E.~Guendelman}{1}, \coauthor{E.~Nissimov}{2},
\coauthor{S.~Pacheva}{2}, \coauthor{M.~Vasihoun}{1}

\address{Department of Physics, Ben-Gurion Univ. of the Negev,
Beer-Sheva 84105, Israel}{1}

\address{Institute of Nuclear Research and Nuclear Energy,
Bulg. Acad. Sci., Sofia 1784, Bulgaria}{2}

\runningheads{E.~Guendelman, E.~Nissimov, S.~Pacheva, M.~Vasihoun}{A New Mechanism of 
Dynamical Spontaneous Breaking of Supersymmetry}

\received{}


\begin{Abstract}
We present a qualitatively new mechanism for dynamical spontaneous breakdown of 
supersymmetry. Specifically, we construct a modified formulation of standard minimal 
$N=1$ supergravity. The modification is based on an idea worked out in detail in 
previous publications by some of us, where we proposed a new formulation of 
(non-supersymmetric) gravity theories employing an alternative volume form 
(volume element, or generally-covariant integration measure) in the pertinent 
Lagrangian action, defined in terms of auxiliary (pure-gauge) fields instead of the 
standard Riemannian metric volume form. Invariance under supersymmetry of the new 
modified $N=1$ supergravity action is preserved due to the addition of an appropriate
compensating antisymmetric tensor gauge field. This new formalism naturally triggers
the appearance of a dynamically generated cosmological constant as an arbitrary 
integration constant which signifies a spontaneous (dynamical) breaking of
supersymmetry. Furthermore, applying the same formalism to anti-de Sitter 
supergravity allows us to appropriately choose the above mentioned arbitrary 
integration constant so as to obtain simultaneously a very small positive effective
observable cosmological constant as well as a large physical gravitino mass as 
required by modern cosmological scenarios for slowly expanding universe of today.
\end{Abstract}

\end{start}

\section[]{Introduction -- Gravity-Matter Theories in Terms of Non-Riemannian
Volume-Forms on Space-Time Manifold}
In a series of previous papers \ct{TMT-orig-1}-\ct{TMT-orig-3} (for recent 
developments, see Refs.\ct{TMT-recent-1,TMT-recent-2})
some of us have proposed a new class of generally-covariant (non-supersymmetric)
field theory models including gravity -- called ``two-measure theories''
(TMT), which appear to be promising candidates for resolution of the dark energy 
and dark matter problems, the fifth force problem, etc. The principal idea is to 
employ an alternative volume form (volume element or generally-covariant integration
measure on the space-time manifold) in the pertinent Lagrangian action, which is  
defined in terms of auxiliary pure-gauge fields independent of the standard 
Riemannian volume form in terms of the Riemannian space-time metric. 

To illustrate the formalism let us consider the following type of TMT action
of the form relevant in the present context::
\be
S = \int d^D\!x \,\Phi(B) \Bigl\lb L^{(1)} + 
\frac{\vareps^{\m_1\dots\m_D}}{(D-1)!\sqrt{-g}} \pa_{\m_1} H_{\m_2\ldots\m_D}
\Bigr\rb + \int d^D\!x \sqrt{-g}\, L^{(2)}
\lab{TMT-action-gen}
\ee
with the following notations:
\begin{itemize}
\item
The first term in \rf{TMT-action-gen} contains a non-Riemannian integration
measure density:
\be
\Phi(B) \equiv \frac{1}{D!}\vareps^{\m_1\ldots\m_D}\, \pa_{\m_1} B_{\m_2\ldots\m_D}
\; ,
\lab{Phi-D}
\ee
where $B_{\m_1\ldots\m_{D-1}}$ is a rank $(D-1)$ antisymmetric tensor gauge
field.
\item
The Lagrangians $L^{(1,2)} \equiv \frac{1}{2\k^2} R + L^{(1,2)}_{\rm matter}$
include both standard Einstein-Hilbert gravity action as well as
matter/gauge-field parts. Here $R=g^{\m\n} R_{\m\n}(\G)$ is the scalar curvature 
within the first-order (Palatini) formalism for the Riemannian space-time metric 
$g_{\m\n}$ and $R_{\m\n}(\G)$ is the Ricci tensor in terms of the
independent affine connection $\G^\m_{\l\n}$. In general, $L^{(2)}$ might
contain also higher curvature terms like $R^2$ (cf. Ref.\ct{TMT-recent-2}).
\item
Note that in the first modified-measure term in the action \rf{TMT-action-gen}
we have included an additional term containing another rank $(D-1)$ antisymmetric
tensor gauge field $H_{\m_1\ldots\m_{D-1}}$. Such terms would be purely topological 
(total divergence) ones if included in standard Riemannian integration
measure action terms like the second term with $L^{(2)}$ on the r.h.s. of 
\rf{TMT-action-gen}.
\end{itemize}
In \rf{TMT-action-gen} we have taken linear combination of modified-measure and 
standard Riemannian measure action terms. Recently in \ct{quintess} we have
proposed a more general TMT gravity-matter model defined in terms of {\em two
different} non-Riemannian integration measures, which provides interesting
cosmological implications, in particular, it allows for a unified description of
both an early universe inflation and present day dark energy.

Varying \rf{TMT-action-gen} w.r.t. $H_{\n\k\l}$ and the ``measure'' tensor gauge
field $B_{\n\k\l}$ we get:
\br
\pa_\m \Bigl(\frac{\Phi(B)}{\sqrt{-g}}\Bigr) = 0 \;\; \to \;\;
\frac{\Phi(B)}{\sqrt{-g}} \equiv \chi = {\rm const} \; ,
\lab{chi-const} \\
L^{(1)} + \frac{\vareps^{\m_1\dots\m_D}}{(D-1)!\sqrt{-g}} \pa_{\m_1} H_{\m_2\ldots\m_D}
= M \; ,
\lab{L1-const}
\er
where $\chi$ (ratio of the two measure densities) and $M$ are arbitrary integration 
constants. Performing canonical Hamiltonian analysis of \rf{TMT-action-gen} one can 
show that the above integration constants M and $\chi$ are in fact constrained
a'la Dirac canonical momenta of the auxiliary tensor gauge fields $B$ and $H$.

Now, varying \rf{TMT-action-gen} w.r.t. $g^{\m\n}$ and taking into account 
\rf{chi-const}--\rf{L1-const} we arrive at the following effective Einstein
equations (in the first-order formalism):
\be
R_{\m\n}(\G) - \h g_{\m\n} R + \L_{\rm eff} g_{\m\n} = \k^2 T^{\rm eff}_{\m\n},
\lab{einstein-eff}
\ee
with effective energy-momentum tensor:
\be
T^{\rm eff}_{\m\n} = g_{\m\n} L^{\rm eff}_{\rm matter} -
2\partder{L^{\rm eff}_{\rm matter}}{g^{\m\n}} \quad, \quad
L^{\rm eff}_{\rm matter} \equiv \frac{1}{\chi + 1} 
\Bigl\lb L^{(1)} + L^{(2)}\Bigr\rb \; ,
\lab{T-eff}
\ee
and with a {\em dynamically generated} effective cosmological constant:
\be
\L_{\rm eff} = \frac{\k^2}{\chi + 1}\,\chi M  \; .
\lab{CC-eff}
\ee

\section[]{Modified $N=1$ Supergravity and Dynamical Supersymmetry Breaking}

The ideas and concepts of two-measure gravitational theories 
\ct{TMT-orig-1}-\ct{TMT-recent-2} 
may be combined with those applied to contsruct a theory of strings and branes with
dynamical generation of (variable) string/brane tension \ct{mstring} to consistently
incorporate supersymmetry into the two-measure modification of standard
Einstein gravity. Here for simplicity we will present the modified-measure
construction of $N=1$ supergravity in $D=4$. For a recent account of
modern supergravity theories and notations, see Ref.\ct{freedman-proeyen}.

Let us recall the standard component-field action of $D=4$ (minimal) $N=1$ supergravity:
\br
S_{\rm SG} = \frac{1}{2\k^2} \int d^4 x\, e
\Bigl\lb R(\om,e) - {\bar\psi}_\m \g^{\m\n\l} D_\n \psi_\l \Bigr\rb \; ,
\lab{SG-action} \\
e = \det\Vert e^a_\m \Vert \;\; ,\;\;
R(\om,e) = e^{a\m} e^{b\n} R_{ab\m\n}(\om) \; .
\lab{curv-scalar} \\
R_{ab\m\n}(\om) = \pa_\m \om_{\n ab} - \pa_\n \om_{\m ab}
+  \om_{\m a}^c \om_{\n cb} - \om_{\n a}^c \om_{\m cb} \; .
\lab{curvature} \\
D_\n \psi_\l = \pa_\n \psi_\l + \frac{1}{4}\om_{\n ab}\g^{ab}\psi_\l \;\; ,\;\;
\g^{\m\n\l} = e^\m_a e^\n_b e^\l_c \g^{abc} \; ,
\lab{D-covariant}
\er 
where all objects belong to the first-order ``vierbein'' (frame-bundle) formalism,
\textsl{i.e.}, the vierbeins $e^a_\m$ (describing the graviton) and the 
spin-connection $\om_{\m ab}$ ($SO(1,3)$ gauge field acting on the gravitino
$\psi_\m$) are \textsl{a priori} independent fields (their relation arises
subsequently on-shell); $\g^{ab} \equiv \h \(\g^a \g^b - \g^b \g^a\)$
\textsl{etc.} with $\g^a$ denoting the ordinary Dirac gamma-matrices.
The invariance of the  action \rf{SG-action} under local supersymmetry 
transformations: 
\be
\d_\eps e^a_\m = \h {\bar\vareps}\g^a \psi_\m \;\; ,\;\;
\d_\eps \psi_\m = D_\m \vareps
\lab{local-susy}
\ee
follows from the invariance of the pertinent Lagrangian density up to a
total derivative: 
\be
\d_\eps \Bigl( e \bigl\lb R(\om,e) 
- {\bar\psi}_\m \g^{\m\n\l} D_\n \psi_\l \bigr\rb\Bigr) = 
\pa_\m \bigl\lb e\bigl({\bar\vareps}\z^\m\bigr)\bigr\rb \; ,
\lab{local-susy-L}
\ee
where $\z^\m$ functionally depends on the gravitino field $\psi_\m$.

We now propose a modification of \rf{SG-action} by replacing the standard
generally-covariant measure density $e = \sqrt{-g}$ by the alternative measure
density $\P(B)$ (Eq.\rf{Phi-D} for $D=4$):
\be
\Phi(B) \equiv \frac{1}{3!}\vareps^{\m\n\k\l}\, \pa_\m B_{\n\k\l} \; ,
\lab{Phi-4}
\ee
and we will use the general framework described above in 
\rf{TMT-action-gen}--\rf{CC-eff}. The modified-measure supergravity action reads:
\be
S_{\rm mSG} = \frac{1}{2\k^2} \int d^4 x\, \Phi(B)\,
\Bigl\lb R(\om,e) - {\bar\psi}_\m \g^{\m\n\l} D_\n \psi_\l 
+ \frac{\vareps^{\m\n\k\l}}{3!\,e}\, \pa_\m H_{\n\k\l} \Bigr\rb \; ,
\lab{mSG-action}
\ee
where a new term containing the field-strength of a 3-index antisymmetric tensor 
gauge field $H_{\n\k\l}$ has been added. As already explained above, its inclusion 
in the Lagrangian of the standard supergravity action \rf{SG-action} would yield a
purely topological (total divergence) term. Similar construction has been
previously used in ref.\ct{mstring} to formulate a new version of
Green-Schwarz superstring using an alternative non-Riemannian world-sheet
volume form.

The equations of motion w.r.t. $H_{\n\k\l}$ and $B_{\n\k\l}$ yield as in
\rf{chi-const}--\rf{L1-const}:
\br
\pa_\m \Bigl(\frac{\Phi(B)}{e}\Bigr) = 0 \;\; \to \;\;
\frac{\Phi(B)}{e} \equiv \chi = {\rm const} \; ,
\lab{chi-const-1} \\
R(\om,e) - {\bar\psi}_\m \g^{\m\n\l} D_\n \psi_\l
+ \frac{\vareps^{\m\n\k\l}}{3!\, e}\, \pa_\m H_{\n\k\l} = 2 M \; ,
\lab{L-const}
\er
where $M$ is an arbitrary integration constant.

Now it is straightforward to check that the modified-measure supergravity
action \rf{mSG-action} is invariant under local supersymmetry transformations
\rf{local-susy} supplemented by the transformation laws for $H_{\m\n\l}$ and $\Phi(B)$:
\be
\d_\eps H_{\m\n\l} = - e\,\vareps_{\m\n\l\k}\bigl({\bar\vareps}\z^\k\bigr)
\quad, \quad \d_\eps \Phi(B) = \frac{\Phi(B)}{e}\,\d_\eps e \; ,
\lab{local-susy-H-Phi}
\ee
which algebraically close on-shell, \textsl{i.e.}, when Eq.\rf{chi-const-1} is
imposed.

The role of $H_{\n\k\l}$ in the modified-measure action \rf{mSG-action} is to absorb, 
under local supersymmetry transformation, the total derivative term coming from 
\rf{local-susy-L}, so as to insure local supersymmetry invariance of \rf{mSG-action} -- 
this is a generalization of the formalism used in Ref.\ct{mstring} to write down a
modified-measure extension of the standard Green-Schwarz world-sheet action
of space-time supersymmetric strings. Similar approach has also been
employed in Refs.\ct{nishino-rajpoot-1,nishino-rajpoot-2} in the context of 
$f(R)$ supergravity.

The appearance of the integration constant $M$ represents a {\em dynamically generated 
cosmological constant} in the pertinent gravitational equations of motion and, thus, 
it signifies a {\em new mechanism of spontaneous (dynamical) breaking of supersymmetry}. 
Indeed, varying \rf{mSG-action} w.r.t. $e^a_\m$:
\br
e^{b\n}R^a_{b\m\n} - \h {\bar\psi}_\m \g^{a\n\l} D_\n \psi_\l
+ \h {\bar\psi}_\n \g^{a\n\l} D_\m \psi_\l 
\nonu \\
+ \h {\bar\psi}_\l \g^{a\n\l} D_\n \psi_\m
+ \frac{e^a_\m}{2}\,\frac{\vareps^{\r\n\k\l}}{3!\, e}\, \pa_\r H_{\n\k\l} = 0
\lab{grav-eqs}
\er
and using Eq.\rf{L-const} to replace the last $H$-term on the l.h.s. of \rf{grav-eqs}
we obtain the vierbein analogues of the Einstein equations including a
dynamically generated {\em floating} cosmological constant term $e^a_\m M$
(cf. Eqs.\rf{einstein-eff}-\rf{CC-eff} above):
\br
e^{b\n}R^a_{b\m\n} -\h e^a_\m R(\om,e) +  e^a_\m M = \k^2 T^a_\m \; ,
\lab{einstein-eqs} \\
\nonu \\
\k^2 T^a_\m \equiv \h {\bar\psi}_\m \g^{a\n\l} D_\n \psi_\l
-\h e^a_\m {\bar\psi}_\r \g^{\r\n\l} D_\n \psi_\l
\nonu \\
- \h {\bar\psi}_\n \g^{a\n\l} D_\m \psi_\l
- \h {\bar\psi}_\l \g^{a\n\l} D_\n \psi_\m \; .
\lab{stress-energy}
\er
Let us recall that according to the classic paper \ct{deser-zumino-77} the sole 
presence of a cosmological constant in supergravity, even in the absence of manifest 
mass term for the gravitino, implies that the gravitino is {\em massive},
i.e., it absorbs the Goldstone fermion of spontaneous supersymmetry breakdown --
a supersymmetric Higgs effect.

\section[]{Dynamical Supersymmetry Breaking in Modified Anti-de Sitter
Supergravity -- Large Gravitino Mass versus Small Cosmological Constant}

Let us now start from anti-de Sitter (AdS) 
supergravity (see, \textsl{e.g.} \ct{deser-zumino-77,freedman-proeyen}; 
using the same notations as in \rf{SG-action}):
\br
S_{\rm AdS} = \frac{1}{2\k^2} \int d^4 x\, e
\Bigl\lb R(\om,e) - {\bar\psi}_\m \g^{\m\n\l} D_\n \psi_\l 
- m\,{\bar\psi}_\m \g^{\m\n} \psi_\n - 2 \L_0 \Bigr\rb \; ,  
\lab{AdS-SG-action} \\
m \equiv \frac{1}{L} \quad ,\quad \L_0 \equiv - \frac{3}{L^2} \; .
\lab{mass-CC}
\er
The action \rf{AdS-SG-action} contains additional explicit mass term for the gravitino 
as well as a bare negative cosmological constant $\L_0$ balanced in a precise way
$|\L_0|=3m^2$ so as to maintain local supersymmetry invariance and, in particular, 
according to \ct{deser-zumino-77} -- keeping the physical gravitino mass zero. 

Then, application of the above formalism from Section 2 to the action \rf{AdS-SG-action} 
allows us to construct a modified-measure AdS supergravity in complete
analogy with \rf{mSG-action}:
\br
S_{\rm mod-AdS} = \frac{1}{2\k^2} \int d^4 x\,\Phi(B)
\Bigl\lb R(\om,e) - {\bar\psi}_\m \g^{\m\n\l} D_\n \psi_\l 
\nonu \\
- m\,{\bar\psi}_\m \g^{\m\n} \psi_\n - 2 \L_0 + 
\frac{\vareps^{\m\n\k\l}}{3!\,e}\, \pa_\m H_{\n\k\l}\Bigr\rb \; ,
\lab{mod-AdS-SG-action}
\er
with $\Phi(B)$ as in \rf{Phi-4} and $m,\L_0$ as in \rf{AdS-SG-action}. 

Repeating the same steps as in Section 2 the AdS action \rf{mod-AdS-SG-action}
will trigger dynamical spontaneous supersymmetry breaking resulting in the appearance 
of the dynamically generated floating cosmological constant $M$ which will add
to the bare cosmological constant $\L_0$. Thus, we can achieve via appropriate choice 
of $M \sim |\L_0|$ a {\em very small positive effective observable cosmological constant}:
\be
\L_{\rm eff} = M + \L_0 = M - 3m^2 \ll |\L_0|
\lab{CC-eff-AdS}
\ee
and, simultaneously, a {\em large physical gravitino mass} $m_{\rm eff}$ which
in the case of very small effective cosmological constant \rf{CC-eff-AdS} will
be very close to the bare gravitino mass parameter ~$m$:
\be
m_{\rm eff} \simeq m = \sqrt{\frac{1}{3}|\L_0|} \; ,
\lab{m-grav-AdS}
\ee
since now the background space-time geometry becomes almost flat.
This is precisely what is required by modern cosmological scenarios for slowly 
expanding universe of today \ct{slow-accel-1}-\ct{slow-accel-3}.

\section[]{Conclusions}
\begin{itemize}
\item
Two-measure formalism in gravity/matter theories (employing alternative non-Riemannian 
volume form, {\em i.e.} reparametrization covariant integration measure, on the spacetime 
manifold alongside standard Riemannian volume form) naturally generates a 
\textbf{\em dynamical cosmological constant} as an arbitrary dimensionful 
integration constant.
\item
Within modified-measure minimal $N=1$ supergravity the dynamically generated cosmological 
constant implies spontaneous supersymmetry breaking and mass generation for the gravitino 
(supersymmetric Brout-Englert-Higgs effect).
\item
Within modified-measure anti-de Sitter supergravity we can fine-tune the dynamically 
generated cosmological integration constant in order to achieve simultaneously a 
{\em very small physical observable positive cosmological constant} and a 
{\em very large physical observable gravitino mass} -- a paradigm of modern 
cosmological scenarios for slowly expanding universe of today.
\end{itemize}

\section*{Acknowledgments}
We gratefully acknowledge support of our collaboration through the academic exchange 
agreement between the Ben-Gurion University and the Bulgarian Academy of Sciences.
S.P. has received partial support from COST action MP-1210.


\end{document}